\documentclass[sigconf]{acmart}

\usepackage{booktabs} 

\setcopyright{rightsretained}
\usepackage{capt-of}
\usepackage{enumerate}
\usepackage{enumitem}
\usepackage{color}
\usepackage{tcolorbox}

\acmDOI{10.475/123_4}

\acmISBN{123-4567-24-567/08/06}

\acmConference[SIGIR 2019]{The 42nd International ACM SIGIR conference on Research and Development in Information Retrieval}{July 2019}{Paris, France}
\acmYear{2019}
\copyrightyear{2019}

\begin{document}
\title{Information search in a professional context -- exploring a collection of professional search tasks}

%

\author{Suzan Verberne}
\affiliation{%
  \institution{Leiden University}
}
\email{s.verberne@liacs.leidenuniv.nl}

\author{Jiyin He}
\affiliation{%
  \institution{Signal Media}
}
\email{jiyin.he@signalmedia.co}

\author{Gineke Wiggers}
\affiliation{%
  \institution{Leiden University}
}
\email{g.wiggers@law.leidenuniv.nl}

\author{Tony Russell-Rose}
\affiliation{%
  \institution{UXLabs}
}
\email{tgr@uxlabs.co.uk}

\author{Udo Kruschwitz}
\affiliation{%
  \institution{University of Essex}
}
\email{udo@essex.ac.uk}

\author{Arjen P. de Vries}
\affiliation{%
  \institution{Radboud University}
}
\email{arjen@acm.org}

\renewcommand{\shortauthors}{Suzan Verberne et al.}

\begin{abstract}
Search conducted in a work context is an everyday activity that has been around since long before the Web was invented, yet we still seem to understand little about its general characteristics. With this paper we aim to contribute to a better understanding of this large but rather multi-faceted area of `professional search'. Unlike task-based studies that aim at measuring the effectiveness of search methods, we chose to take a step back by conducting a survey among professional searchers to understand their typical search tasks. By doing so we offer complementary insights into the subject area. We asked our respondents to provide actual search tasks they have worked on, information about how these were conducted and details on how successful they eventually were. We then manually coded the collection of 56 search tasks with task characteristics and relevance criteria, and used the coded dataset for exploration purposes. Despite the relatively small scale of this study, our data provides enough evidence that professional search is indeed very different from Web search in many key respects and that this is a field that offers many avenues for future research.
\end{abstract}

%
%


\maketitle

\section{Introduction}
Professional search is defined as the searching carried out by experts for work purposes~\cite{koster2009challenges,russell2018information}. According to the literature, three specific characteristics of professional search differentiate it from web search: (a) The search tasks are \textbf{complex and specific}~\cite{Mason2006,morville2010search}; (b) The professional searcher is an \textbf{expert} in the search domain~\cite{hjorland2015classical};
(c) Professionals have \textbf{exploratory search needs}~\cite{he2013characterizing} that require multiple queries and include browsing and analysing multiple documents~\cite{makri2008investigating}.\footnote{Note that point (b) differentiates professional search from enterprise search \cite{INR-053}.\label{fn:es}}

Search conducted in a work context is an everyday activity that has been around since long before the Web was invented. However, professional search has gained much less attention from the academic IR community than web search. As a result, many aspects of professional search are still unknown~\cite{verberne2018first,DBLP:conf/desires/Lewis18}. Therefore even small, qualitative data sets can be valuable. They allow us to learn, case by case, how to approach difficult search tasks in realistic settings. 

With this paper we contribute to a better understanding of professional search. Unlike system-oriented studies that aim at measuring the effectiveness of search methods for a particular task, we decided to take a step back by conducting a survey among professional searchers to understand their typical search tasks. We asked our respondents to provide actual search tasks they have worked on, information about how these were conducted and details on how successful they eventually were. We then manually coded the collection of 56 search tasks with task characteristics and relevance criteria, and used the coded dataset for exploration purposes. 
We address the following research questions:
\begin{enumerate}[label=RQ\arabic*]
\item To what extent are the characteristics of professional search (a)--(c) reflected by the data acquired in our survey?
\item Are these characteristics sufficiently pronounced to justify treating professional search as a separate search genre? 
\item Are the needs, goals and behaviours of professional searchers sufficiently homogeneous and consistent to justify viewing `professional search' as a coherent, single field of enquiry? 
\end{enumerate}

\section{Background and related work}
\label{sec:relwork}
\emph{Collections of professional search tasks.} Most collections of search tasks have been created in the context of TREC. A few of the relevant tracks for professional search are the total recall track, the enterprise track, and the legal track. TREC collections have been designed for evaluation purposes; they have shown to be indispensable for system comparisons and benchmarking. 
To the contrary, our collection is not meant for benchmarking purposes. The collection comprises user-created descriptions of a work task, created after completion of the task. 
Although our work does not contribute a test collection (it has no relevance assessments), the topics in our collection were collected similar to those in the iSearch collection~\cite{lykke2010isearch}, where the searchers themselves (experts) formulated topics that include fields for work task context, information need, and ideal answer. 

\emph{Coding schemes for search tasks.} Search tasks can be coded in a variety of ways. In early work, Bates \cite{bates1979information} identified a set of 29 search `tactics' which she organised into four broad categories of information seeking behaviour. 
Ellis and colleagues \cite{ellis1997modelling} developed a model consisting of a number of broad information seeking behaviours, noting also that it is possible to display more than one behaviour at any given time. Makri et al \cite{makri2008investigating} extended this work, focusing on the information behaviours observed within the legal profession. More recently, Russell-Rose et al \cite{russell2011taxonomy} used a grounded-theory approach to identify a taxonomy of information behaviours derived from a corpus of enterprise search tasks.

In many of these previous studies of information seeking behaviour, interview transcripts have served as the primary data source, offering an indirect, verbal account of end user information behaviours. By contrast, the data source used in this study represents a self-reported account of information behaviour, generated directly by end users (albeit retrospectively). 

\emph{Relevance criteria.} 
In 1998, Rieh and Belkin~\cite{rieh1998understanding} 
identified seven different factors of information quality: source, content, format, presentation, currency, accuracy, and speed of loading, and two different levels of source authority: individual and institutional. Savolainen and Kari~\cite{savolainen2006user} found in an exploratory study that specificity, topicality, familiarity, and variety were the four most mentioned criteria in user-formulated relevance judgments, but there was a high number of individual criteria mentioned by the participants. We include the relevance factors from these prior works in the scheme for coding the relevance criteria mentioned by our respondents.

\section{Data}

\subsection{Collecting search tasks}
\begin{figure}\begin{tcolorbox} You could help us collecting example search tasks. Please describe one search task that you have undertaken recently. Please be as specific as possible (not: ``trying to find papers'', but ``trying to find papers that referenced Ingwersen and J\"{a}rvelin (2005)''). \begin{itemize} \item What was the goal of the search? \item Describe which actions you took to search (which search engine, queries, metadata filters) \item Describe how you judged the relevance of the found information. What were the relevance criteria? \item What was the outcome of the search? (did you find what you were looking for; how satisfied were you?) \end{itemize}\end{tcolorbox}\vspace{-0.4cm} \caption{Instructions given to respondents of our survey} \vspace{-0.4cm}\label{fig:instructions}\end{figure}

The data was collected by surveying a non-representative sample of professional searchers.
We defined the target group of the survey as ``everyone who regularly performs complex search tasks at work in environments other than general web search''. This included information specialists in various domains, but also librarians, scientists, lawyers, and other knowledge worker professions. 
We distributed the survey in our own networks (e-mail, LinkedIn, and other social media), and through a number of professional mailing lists and newsletters that are distributed among information specialists and librarians in several domains. The instructions given to the respondents are shown in Figure~\ref{fig:instructions}. We also asked them about their field of expertise, their age and the number of years of professional experience.

\subsection{Manual coding}

\begin{figure}[t]\begin{tcolorbox}
We manually coded the search tasks according to the following dimensions: \begin{itemize}[leftmargin=*] \item Topic is well defined; keep in the set (yes/no/not sure) \item Topic domain (Computer science / Humanities / Legal / Medical / Other) \item  Search for self or other (Self / Other / Unknown) \item Number of search systems/databases mentioned \item Satisfaction score (2: good; 1: sufficient; 0: undefined; -1: negative) \item Relevance criteria (Document type; Expertise level; External relevance; Language; Source reliability; Publication date; Quality; Sensitivity/Recall; Specificity/Precision; Topical relevance\\ (True or false, multiple options possible) \end{itemize} \end{tcolorbox} \vspace{-0.4cm}\caption{Coding scheme used by the authors to code the collected search tasks. We refer to the first 5 as `the main task characteristics'.}\vspace{-0.4cm} \label{fig:codingscheme}\end{figure}

In order to aggregate more structured information about the collected search tasks and quantify characteristics, we manually coded the set of search tasks. 
Analyzing the answers to the survey questions enabled us to transform the natural language descriptions of search tasks into quantitative measures such as the number of search systems used for the task, and the degree of satisfaction with the results obtained.
An additional purpose of the manual coding was to remove under-specified search tasks, such as ``search for systematic review'' or ``Trying to find papers and citations''. 

For defining the coding scheme, we followed a grounded theory approach: we based our coding scheme on existing schemes for search task coding and relevance criteria (see Section~\ref{sec:relwork}), making adaptations to the categories to fit our data. For example, for topic domain we include the domains that occur in our data. 

Our coding process was as follows: The first round of coding on a small sample of the topics was done by the first author. Based on the findings the initial coding scheme was defined. The second round of coding on a small sample of the topics was done by one of the other authors, using the initial coding scheme. The differences between the codings of the two coders were discussed and the coding scheme was revised where needed. The third round of coding was done by all authors, on the complete set. Each search task was coded by exactly three coders. The resulting coding scheme is summarized in Figure~\ref{fig:codingscheme}.\footnote{The actual coding scheme included more elaborate explanations for each dimension, and will be shared online (URL excluded for anonymous peer review).}

\subsection{Merging the annotated sets}
The codings were merged into one set of coded search tasks. If at least one of the three coders voted to exclude the topic from the data (first question in the coding scheme), the topic was excluded. Codings for the remaining topics were combined as follows: (1) If at least two coders agree on the value for the dimension, that value was assigned (all disagreements for the -- binary -- relevance criteria were solved this way); (2) if all three coders selected a different value, we assigned the median in case of numeric values, or took coder 1's (the first author's) label in case of nominal values. The latter happened only for one item in our data.

\section{Results}

71 respondents submitted a search task, out of which 15 search tasks were removed because they were not sufficiently specific. Thus, the resulting collection of professional search tasks, analysed below, contains 56 topics. 

\subsection{Statistics on the manual coding}

\begin{table}[t]
\caption{Agreement statistics on the manually coded task characteristics. Underlined $\kappa$ values indicate substantial or near-perfect agreement ($\kappa >= 0.6$). Italic $\kappa$ values indicate moderate agreement ($0.4 <= \kappa < 0.6$).}\vspace{-0.4cm}
{\small
\begin{tabular}{lp{1.3cm}r}
\hline
Variable & Absolute agreement & Cohen's $\kappa$\\
\hline
Topic domain & 84\% & \underline{0.70}\\
Number of search systems mentioned & 91\% & \underline{0.89}\\
Satisfaction score  & 88\% & \underline{0.82}\\
Search for self or other & 71\% & \emph{0.56}\\
\hline
\end{tabular}}
\label{tab:kappa}
\end{table}

We measured the inter-rater agreement for each dimension in the coding scheme using Cohen's $\kappa$ on the three pairs of annotators. We report mean $\kappa$ scores over the pairs. In the case of numeric variables we computed weighted $\kappa$, in which exact agreement is counted as 1, a difference of 1 is counted as 2/3, a difference of 2 is counted as 1/3 and a difference larger than 2 is counted as 0. The agreement statistics for the main task characteristics are in Table~\ref{tab:kappa}. The $\kappa$  values indicate that agreement is substantial or near-perfect for three of the four task characteristics. The last one (search for self or other) has a moderate agreement. This characteristic is often difficult to judge from the description by the searcher. The moderate agreement is therefore caused by the occurrence of `Unknown' as label for this dimension, occurring in 23 out of 56 search tasks.

\subsection{Statistics on the respondents}

The most represented age group is 46--55 (36\%), followed by 36--45 (27\%) and 56--65 (20\%). 39\% of the respondents has over 20 years of professional experience; 34\% has between 11 and 20 years of professional experience. The majority of respondents (36) listed Library and Information Science (LIS) as field of expertise, followed by Healthcare/biomedical (20) and Computer science (7).\footnote{The sum is more than 56 because respondents were allowed to select more than one field.}

\subsection{Statistics on the search tasks}

\begin{figure}[t]
\includegraphics[width=9cm]{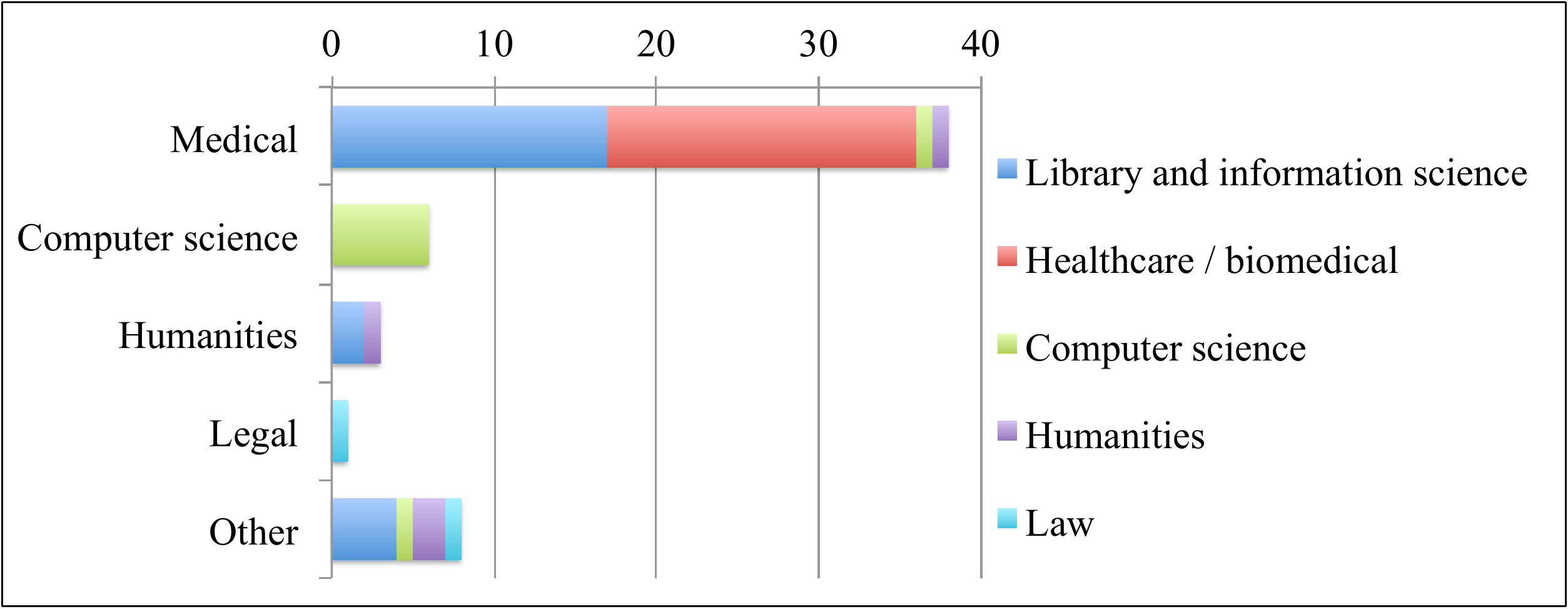}
\vspace{-0.4cm}
\caption{Frequencies of occurrence of topic domains of the search tasks. Each bar represents one topic domain; the colors within the bars represent the first field of expertise listed by the respondent for the search task.}
\label{fig:domain}
\end{figure}

\emph{Topic domain.} Figure~\ref{fig:domain} shows the frequencies of topic domains coded for the search tasks. The majority of search tasks take place in the medical domain. A substantial part of those tasks was submitted by a LIS expert. Interestingly, one task was coded as medical while the respondent listed Humanities as field of expertise. This was a task submitted by a Communications adviser who was researching the availability of online information for cancer patients.

\emph{Search for self or other.} For as few as 11 search tasks, it was explicitly mentioned that the searcher searched for themselves. For 22 search tasks, the respondent explicitly mentioned searching for someone else. In the remaining 23 cases it was not clear to us whether the search task was for the searcher themselves or for an external requester. The respondents who search for an external requester largely listed Library and Information Science (LIS) as their field of expertise. If we assume that all the LIS experts search for others, 17 of the unknowns can additionally be classified as `Search for other', which makes the majority even larger: 40 out of 56 search tasks were executed for an external requester.

\emph{Number of search systems used.} The median number of search systems/databases used to complete one search task was 3 (mean: 3.1; standard deviation: 2.5). An example response to the question ``Describe which actions you took to search'' with 3 databases mentioned is: ``Used NICE HDAS to search Embase, Medline and PubMed. Used a mixture of thesaurus terms and keywords and combined using Boolean operators''.

\emph{Satisfaction with the results.} The main satisfaction score over the search tasks was 1.1, indicating `sufficient result'. For 6 search tasks, the respondent was negative about the result (score -1); for 25 search tasks the result was good (score 2).

\begin{figure}[t]
\includegraphics[width=9cm]{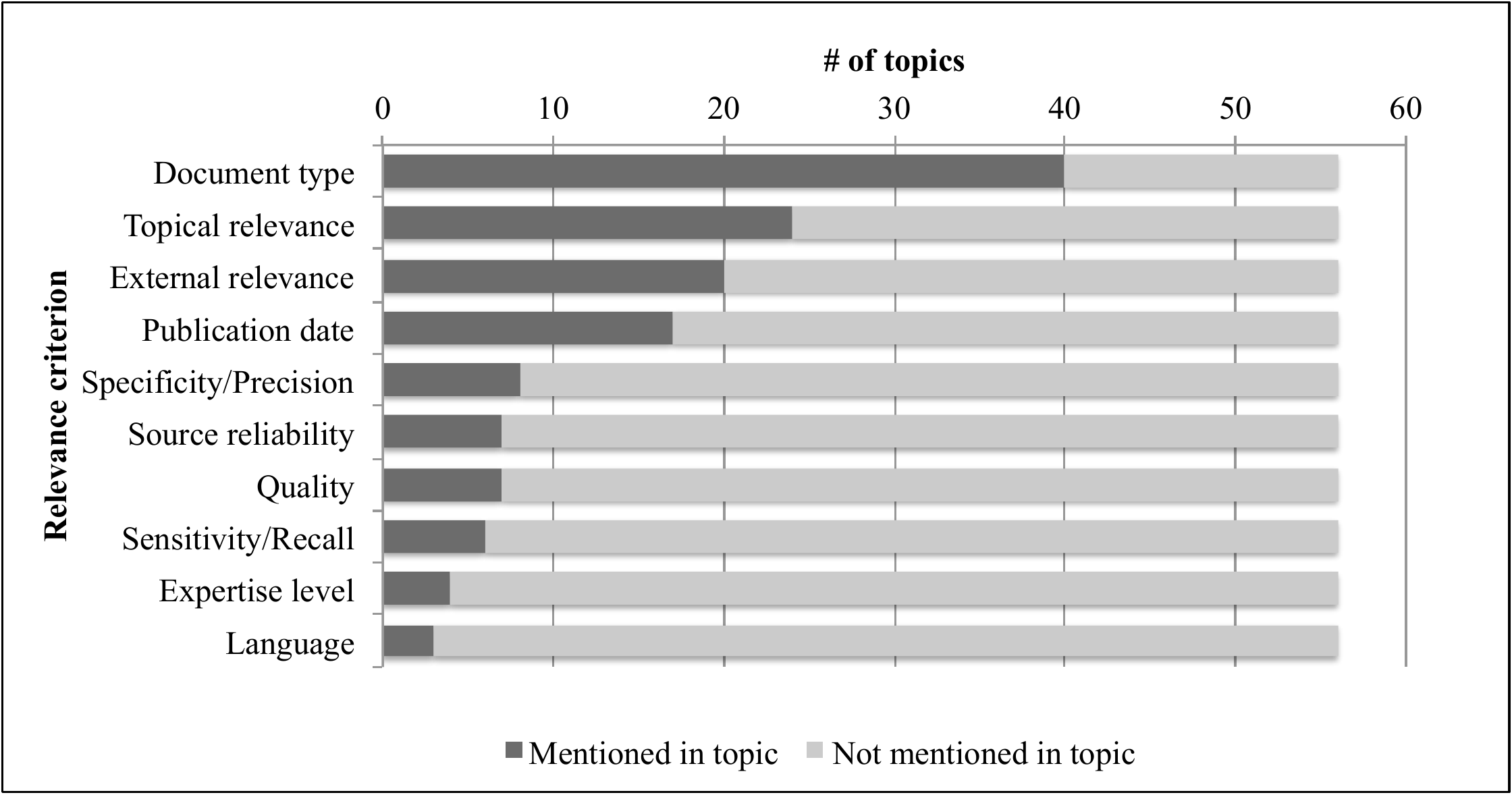}
\vspace{-0.4cm}
\captionof{figure}{Number of search tasks in which each of the relevance criteria is mentioned by the respondent}
\vspace{-0.4cm}
\label{fig:criteria}
\end{figure}


\emph{Relevance criteria.} Figure~\ref{fig:criteria} shows the frequencies of relevance criteria mentioned in the search tasks.  
Two thirds of the relevance descriptions mention the document type. Most of the time this was a (scientific) article. Topical relevance was mentioned in 24 topics, often described in terms of `aboutness' (e.g. ``Articles about patient call, alert, alarm systems that used phones to call nurses''). A relevance criterion that is typical for information specialists is `external relevance', denoting that someone else than the searcher themselves -- the requester of the information -- determines the relevance of the result. This mostly is evidenced by responses such as ``Passed on all relevant papers for expert requester to sift''. The fourth most frequently mentioned relevance criterion is publication date, mostly a mention of `recent papers'.

\section{Discussion}

From our collection of professional search tasks we can distill the following findings about characteristics of professional search (a)--(c) listed in the introduction:

\emph{(a) The search tasks are complex and specific.} Complex search tasks are tasks that need multiple steps to be completed, by collecting information from multiple sources~\cite{he2013characterizing}. From our data, 
it is clear that multiple sources were needed to complete the tasks. The median number of search systems/databases mentioned in one search task was 3 and there were search tasks with over 8 search systems required to complete the task. Some of the respondents explicitly show the complexity of their queries; 9 search tasks mention or show the use of Boolean operators. With respect to the specificity of the search tasks, we see that the tasks are clearly domain-specific, and addressing highly specialized topics. 

\emph{(b) The professional searcher is an expert in the search domain.} In the set of responses to our survey this is not necessarily the case. Figure~\ref{fig:domain} showed that a large portion of the search tasks in the medical domain was completed by LIS experts. The recruitment of respondents explicitly included information specialists and librarians. We did not foresee that this target group would make up the majority of our respondents. This aspect of the survey does not change the other characteristics of the search tasks. In addition, LIS experts are experts in information search, which makes them different from users in the context of enterprise search (see footnote \ref{fn:es}), in which any staff member is a potential user. 


\emph{(c) Professionals have exploratory search needs that require multiple queries and include browsing and analysing multiple documents.} As said above, multiple sources are used in many of the search tasks, probably partly caused by the importance of completeness of the result set (systematic review tasks, which are inherently recall-oriented). We do not have quantified results on the exploratory nature of the search tasks, but most of the topics seem well-defined and specific rather than exploratory.

\section{Conclusions}


\emph{RQ1. To what extent are the characteristics of professional search (a)--(c) reflected by the data acquired in our survey?}
The results of our survey reflect that the search tasks in professional search are complex and highly specific, but not necessarily exploratory. The results also show that not every professional searcher is an expert in the search domain; they can also be LIS experts.

\emph{RQ2. Are these characteristics sufficiently pronounced to justify treating professional search as a separate search genre?}
The characteristics that we found confirm the differences between professional search and web search mentioned in the literature. These have implications for the design of professional search systems. First, the finding that many professional searchers search for others means that the searcher may not be in a good position to assess the relevance of results. For that reason it might be a good idea to provide additional information in the interface of the IR system based on possible relevance criteria (e.g. publication date, expertise level)~\cite{Wiggers2018}, to aid the user in creating the (short)list for the client. 
Second, the complex information needs of professional searchers, for example in systematic reviews, means that professional searchers are searching for information spread across multiple documents. This means that there is no one particular document that best provides the information, and that there is no clear requirement what document should be ranked first. The user interface of a professional search system should ideally be adapted for this characteristic, and show a result set covering a diverse set of aspects to the information need.

\emph{RQ3. Are the needs, goals and behaviours of professional searchers sufficiently homogeneous and consistent to justify viewing `professional search' as a coherent, single field of enquiry?}

Based on the data of this research, it seems that the same characteristics of professional search apply to professional searchers from all the domains that participated in our study. Further research is required before it can be determined whether all groups can be treated as one field of enquiry for these purposes. A prominent finding from our survey is the evidence of multiple sources being used -- and that the tools do not support this that well. As a result, many different search engines are used to search the relevant sources.



\bibliographystyle{ACM-Reference-Format}
\bibliography{profsearch}

\end{document}